\documentclass{aa} 
\usepackage{graphicx}
\newenvironment{deluxetable}[1]
{\begin{table}\caption[]{\ptcaption}
\begin{flushleft}\edef\tableformat{\string#1}\ptcolsep
\begin{tabular}{\tableformat}}{\noalign{\smallskip}\hline
\noalign{\medskip}\end{tabular}
\\\ptcomments\medskip\ptrefs\end{flushleft} 
\end{table}}

\newenvironment{deluxetable*}[1]
{\begin{table*}\caption[]{\ptcaption}
\begin{flushleft}\edef\tableformat{\string#1}\ptcolsep
\begin{tabular}{\tableformat}}{\noalign{\smallskip}\hline
\noalign{\medskip}\end{tabular}
\\\ptcomments\medskip\ptrefs\end{flushleft}
\end{table*}}

 \def\ptcolsep{\relax}
\def\tablecaption#1{\gdef\ptcaption{#1}} \def\ptcaption{\relax}

 \def\ptcomments{\relax}

 \def\ptrefs{\relax}

\newcommand{\tablehead}[1]{\hline\noalign{\smallskip}#1\\}
\newcommand{\colhead}[1]{\multicolumn{1}{c}{#1}}
\newcommand{\startdata}{\noalign{\smallskip}\hline\noalign{\smallskip}}

\newcommand{\tablewidth}[1]{\typeout
{----- tablewidth not implemented with A\&A ----------------------}}{}

\newcommand{\realfiguremid}[3]{
              \begin{figure*}
              \includegraphics[width=11.5cm]{#1}\hfill
              \parbox[b]{55mm}{
              \caption{#2}\label{#3}}
              \end{figure*}}

\usepackage[dvips]{color}

\newcommand{\abbrev}[1]{#1}      %%  plain, final
\newcommand{\di}{dIrr} 
\newcommand{\psf}{PSF} 
 
\newcommand{\agb}{{\sc AGB}}
\newcommand{\rgb}{{\sc RGB}} 
\newcommand{\grotto}{GR\,8}
\newcommand{\allframe}{{\sc allframe}}
\newcommand{\etal}{{et al.}\ }

\newcommand{\daophot}{{\sc daophot ii}}
\newcommand{\allstar}{{\sc allstar}}
\newcommand{\cmd}{{CMD}}

\newcommand{\ism}{{ISM}}

\newcommand{\hi}{\hbox{H{\sc i}}}
\newcommand{\hii}{\hbox{H{\sc ii}}}
\newcommand{\oiii}{\hbox{O{\sc iii}}}
\newcommand{\msol}{$M_{\odot}$}

\newcommand{\kms}{km s$^{\rm -1}$}

\newcommand{\gtrsim}{\ga}  
\begin{document}

%% \thesaurus{11.06.2; 11.09.1 SagDIG; 11.12.1; 11.19.5; 11.19.6}

\title{
The Sagittarius dwarf irregular galaxy: metallicity and stellar 
populations
%A key galaxy in the Local Group. Reassessment of the metallicity 
%and stellar content of SagDIG
\thanks{Based on data collected at the European Southern Observatory,
         La Silla, Chile, prop. No. 63.N-0024}}
%\subtitle{}

%IS:
% considerato che il 'movimento' su SagDIG l'ho iniziato io, mi
% pare di essere finito un po' indietro nella lista degli autori...
%
\author {
Y. Momany \inst{1,2}
\and
E. V. Held  \inst{3} 
\and 
I. Saviane \inst{4}
\and
L. Rizzi \inst{3,2}
}
\offprints {Y. Momany}

\institute{
European Southern Observatory, Karl-Schwarschild-Str. 2, D-85748
Garching, Germany
\and
Dipartimento di Astronomia, Universit\`a di Padova,
vicolo dell'Osservatorio 2, I--35122 Padova, Italy \\
\email{momany@pd.astro.it}
\and
Osservatorio Astronomico di Padova, 
vicolo dell'Osservatorio 5, I--35122 Padova, Italy
\email{(held,rizzi)@pd.astro.it}
\and 
European Southern Observatory, Casilla 19001, Santiago 19, Chile
\email{isaviane@eso.org}
}

\date {}
\titlerunning{The SagDIG dwarf : metallicity and stellar populations}
%%%\maketitle

%+++++++++++++++++++++++++++++++++++++++++++++++++++
\abstract{We present deep $BVI$ observations of the dwarf
irregular galaxy UKS1927-177 in Sagittarius (SagDIG). Statistically
cleaned $V$, $(B-I)$ color-magnitude diagrams 
clearly display the key evolutionary features in this galaxy.
Previously detected C stars are located in the color-magnitude
diagrams and shown to be variable, thus confirming the presence of a
significant upper-AGB intermediate age population. A group of likely
red supergiants is also identified, whose magnitude and
color is consistent with a 30 Myr old burst of star formation.
The observed colors of both blue and red stars in SagDIG are best
explained by introducing a differential reddening scenario in which
internal dust extinction affects the star forming regions.
Adopting a low reddening for the red giants, $E(B-V) = 0.07 \pm 0.02$,
gives [Fe/H]=$-2.1 \pm 0.2$ for the mean stellar metallicity, 
a value consistent with the [O/H] abundance measured in the
\hii\ regions. This revised metallicity, which is in accord with the trend of
metallicity against luminosity for dwarf irregular galaxies, is indicative 
of a ``normal'', although metal-poor, \di\ galaxy.
% The distance measured from the tip of the red giant branch is
% $25.14\pm0.18$, in agreement with previous authors.
%
A quantitative description is given of the spatial distribution of
stars in different age intervals, in comparison with the distribution
of the neutral hydrogen.  We find that the youngest stars are located
near the major peaks of emission on the \hi\ shell, whereas the red
giants and intermediate-age C stars define an extended halo or disk
with scale length comparable to the size of the hydrogen cloud.  The
relationship between the distribution of \ism\ and star formation is 
briefly discussed.
\keywords{Galaxies: fundamental parameters -- SagDIG
-- Local Group -- stellar populations -- star formation history.}
}  %%% \end{abstract}

%interesting Star Formation History (\sfh) in which the galaxy
%experienced a quiescent phase a 100 Myr ago, at which it stopped
%forming stars. While the most recent burst happened 31.6 Myr with a
%declining Star Formation Rate (\sfr) down to 5 Myr, characterized by a
%sudden drop around 13 Myr.  The most recent star formation is located
%in a clump that is surrounded on three sides by \hi\ peaks while
%on the fourth lies the \hi\ hole.  A similar neutral hydrogen
%structure (of holes and shells) has been seen in another \di, Sextans
%A (Van Dyk \etal
%\cite{vandyk98}).  We show evidence of the presence of an older (HeB)
%population {\em within} the \hi\ hole, thus consistent with their
%scenario of an expanding front of star formation outward from the
%center.

\maketitle

%+++++++++++++++++++++++++++++++++++++++++++++++++++
\section{Introduction}
\label{s_intro}

% IS:
% sommario vecchi lavori OK; ma quali sono gli shortcomings, e perche'
%  e' necessaria questa nuova ricerca?
% Appena mi sono riletto i due vecchi paper, ci scrivo qualcosa
%
The Sagittarius dwarf irregular (\abbrev{\di}) galaxy, also known as
SagDIG or UKS~1927-177, is a quite difficult object to study because
of its low Galactic latitude and consequent high foreground
contamination.  First reported by 
Cesarsky \etal (\cite{cesarsky77}) on ESO Schmidt photographic plates,
the galaxy was studied by Longmore \etal (\cite{longmore78}), who
derived a total luminosity $M_B=-10.5$. The first CCD study of the
resolved star content in SagDig was that of Cook (\cite{cook87};
hereafter C87), who obtained both intermediate-band and broad-band
photometry. 
%
% From the $(V-I)$, $I$ color-magnitude diagram
% (\abbrev{\cmd}), he derived a distance modulus in the range 25--26 by
% averaging the results of four distance indicators, namely the
% brightest blue and red stars, the tip of the red giant branch
% (\abbrev{\rgb}), and the carbon star luminosity function.
% %
% % Karachentsev \& Tikhonov (\cite{kara93}), and Rozanski \&
% % Rowan-Robinson (\cite{roza94}), showed clearly that one can only 
% % give a first estimate of the true distance if relying on the brightest 3
% % stars method; they estimated a limit of accuracy of 0.5 mag at best
% %
% This remained for a long time the only study of the stellar population
% in SagDIG. 
%
%
Recently, two new investigations have improved our
knowledge of the distance and stellar content of this galaxy
(Karachentsev \etal \cite{kara99}; Lee \& Kim \cite{lee00}; hereafter
KAM99 and LK00 respectively). A new distance modulus was derived from
the tip of the \rgb, equal to $(m-M)_0=25.13$ in KAM99 and 
$(m-M)_0=25.36$ in LK00, respectively. 
This distance indicates that SagDIG is a member of the Local Group
(LG), confirming the evidence from its negative radial velocity and
position in the $V_{\odot}$ $vs.$ $\cos{\theta}$ diagram (van den
Bergh \cite{vandenberg94}; Pritchet \& van den Bergh
\cite{pritchet99}).
%
%The integrated luminosity turned out to be $M_V = XXX$, considerably
%brighter than the first measurements had suggested.  
Both studies, while adopting different estimates for the reddening,
concluded that SagDIG has the lowest metallicity among the LG star
forming dwarf galaxies (KAM99 estimated [Fe/H]$\sim-2.45$, while 
LK00 give [Fe/H] in the range $-2.8$ to $-2.4$; both estimates 
are based on an extrapolation of the calibration relation for 
Galactic globular clusters). 

%-- THE INTERSTELLAR MEDIUM:
The cold and warm interstellar medium (\abbrev{\ism}) of SagDIG and
its kinematics have been the subject of several investigations.
Skillman \etal (\cite{stm89}) obtained optical spectrophotometry of
the most luminous \hii\ regions, and estimated a [O/H] abundance
$\sim3\%$ of the solar value.
%--- (O/H) for         SagDIG is = 2.6 \pm 1.3 \times 10^{-5}
%--- (O/$H_{\odot}$) Solar is = 8.3 \times         10^{-4}  
%
New measurements of the O and N abundance in the \hii\ regions of
SagDIG are presented in a companion paper (Saviane et
al. \cite{savi+01}).  The new estimate, $12+\mbox{log(O/H)}=7.23\pm
0.20$, is by 0.2 dex more metal-poor than found by Skillman et
al. (\cite{stm89}).
The photometric properties of the \hii\ regions were investigated by
Strobel \etal (\cite{strobel91}), who detected 3 regions in their
$2\farcm5$ square field. 
% The lack of stellar photometry prevented any
% identification of the exciting stars inside these \hii\ regions.

%--
High-resolution, high sensitivity VLA observations of the SagDIG \hi\
content have been obtained by Young \& Lo (\cite{young97}) (see also
Lo et al. \cite{lo+sar}).  About $1.2 \times 10^7$ \msol\ of H$+$He
have been estimated (using the new distance from KAM99 and the \hi\
mass from Young \& Lo),
%-- 8.6e6 HI x 1.4 (to H+He)
distributed in an almost symmetric ring likely produced by the
combined effects of stellar winds and supernovae.
SagDIG thus appears to have a high mass fraction in the form of
neutral gas, $M_{H I}/L_B= 1.6$, a value quite typical of the \hi\ 
content in dwarf irregular galaxies (see Mateo \cite{mateo98}).
%
% The \hi\ extends out to 3 Kpc from the galaxy center (about 3 times
% the size of the galaxy on the Palomar Sky Survey).  The \hi\ gas does
% not show rotation, and appears to derive support against gravity from
% random motions. Young \& Lo (\cite{young97}) showed that the \ism\ of
% SagDIG comprises a broad, warm component ($\sigma=10$ \kms) and a
% narrow, cold one ($\sigma = 5$ \kms), with the warm component
% distributed throughout the galaxy while the cold one is concentrated
% into a small number of clumps of $\sim8\times10^5$ \msol. A
% prominent clump is nearly coincident with the biggest \hii\
% region. The cold \hi\ phase may be more directly associated to the
% star formation process.
% %
% The lack of rotation of SagDIG makes a determination of the total mass
% difficult.  Assuming a radius of 1.6 Kpc and a velocity dispersion of
% 10 \kms, Young \& Lo (\cite{young97}) obtained a virial mass of
% 1$\times10^8$ \msol. However, the structure of the \hi\ ring suggests
% that a virial model is probably not appropriate.

%----- OUR WORK:
Because of its high gas content, low luminosity, and especially its
claimed very low metallicity, SagDIG may be a clue to the origin and
evolution of dwarf galaxies.
Although located at the border of the Local Group
it is still close enough to allow us to study both its stellar
populations and the gaseous component, and to compare their physical
properties. In particular, a sound knowledge of the metallicity of
SagDIG is important to constrain the luminosity-metallicity relation
for {\di}s at the metal-poor end, and to trace the metal enrichment
history in dwarf galaxies.
All this motivated an independent study of the metallicity, distance,
and stellar content of SagDIG. The large baseline provided by the
$(B-I)$ color, together with an analysis based on statistical
subtraction of the Galactic foreground, allowed us to improve the
discrimination of young and old stellar populations in the
color-magnitude diagram (CMD). By accounting for the differential
effects of internal dust extinction, we revise upward the [Fe/H]
estimate and show that the metal abundances of the red stars and the
\ism\ can be easily reconciled.
%----- PLAN OF THE PAPER:

The plan of the paper is as follows.
In Sect.~\ref{s_observ} we present the data reduction and calibration.
Section~\ref{s_cmd} presents color-magnitude diagram in different
colors, using a statistical correction of the foreground
contamination. The CMD location of C stars is also discussed.
In Sect.~\ref{s_basic} we rederive the distance and metallicity of
SagDIG by assuming a different reddening for the young and old
populations.
The properties and different spatial distributions of the blue and red
stellar populations are quantified and discussed in
Sect.~\ref{s_stelpop}, where 
the surface density of young stars in different
age intervals is compared with the projected distribution of
neutral hydrogen. As a result of the revised metallicity, SagDIG is
shown to fit well the known luminosity-abundance trend of dwarf
irregular galaxies.
Our results are summarized in Sect.~\ref{s_summary}.

%++++++++++++++++++++++++++++++++++++++++++++++++++++++++++++++++++
\section{Observations and data reduction}
\label{s_observ}

%--------------------------------------
%\subsection{Observations}

Observations of the Sagittarius dwarf irregular were obtained in two runs. 
SagDIG was first observed on the photometric night of Sept.~8, 1996, using 
{\sc efosc2} at the ESO/MPI~2.2m telescope. 
The {\sc efosc2} camera was equipped with a $2048\times2048$ Loral CCD, 
with pixel size of 0\farcs26. A readout window was employed to discard the 
vignetted edges of the CCD, yielding a useful area of $6\farcm 9 \times 
6\farcm 9$. Since the tidal radius of SagDig is only  $\sim 1\farcm7$ (Mateo 
\cite{mateo98}), no observations of a separate control field were needed.
A detailed account of the observations and reduction for this run
can be found in our study of the Phoenix dSph/dI galaxy 
(Held \etal \cite{held99}). 
The galaxy was re-observed in better seeing conditions on the 
night of 30 Sept. 1999 with the EMMI multi-mode instrument on the 
ESO NTT telescope. The sky was marginally photometric with 
occasionally some thin cirrus. The total exposure times were 1$^h$\ in $V$, 
$20^m$\ in $B$, and $40^m$\ in $I$.  The $V$ and $I$ data were obtained 
using a 2048$\times$2048 Tektronix CCD at the red arm of EMMI, 
with pixel size of 0.24 $\mu m$ (0\farcs27) and a field of view 
of $9\farcm2 \times 9\farcm2$. 
The $B$ images were taken at the blue arm of EMMI, with a 
$1024 \times 1024$ Tektronix CCD with a projected pixel size of 
0\farcs37 and a smaller field-of-view, $6\farcm2 \times 6\farcm2$. 
%
% Both detectors were read in fast mode, giving a readout noise 7.3
% (4.0) $e^-$ pixel$^{-1}$ and a gain of 2.76 (2.16) $e^-$ ADU$^{-1}$ at
% the blue (red) arm, respectively.
%
The observing log for both runs is given in Tab.~\ref{t_obsjou}.

%------------------------------------------------------------------->
\tablecaption{The journal of observations \label{t_obsjou}}
\begin{deluxetable}{llcrcc}
\tablehead{
\colhead{Night} &
\colhead{Tel.} &
\colhead{Filter} &
\colhead{$t_{\rm exp}{[s]}$} &
\colhead{$X$} &
\colhead{FWHM} }
\startdata
Sept.~8, 1996 & 2.2m  &  $B$ & 2$\times$900 & 1.02 & $1\farcs3$    \\ 
Sept.~8, 1996 & 2.2m  &  $V$ & 2$\times$600 & 1.05 & $1\farcs5$    \\ 
Sept.~8, 1996 & 2.2m  &  $I$  & 2$\times$900 & 1.03 &   $1\farcs4$  \\ 
Sept.~30, 1999 & NTT  &  $B$ & 4$\times$300   & 1.08 & $1\farcs0$    \\ 
Sept.~30, 1999 & NTT  &  $V$ & 12$\times$300 & 1.08 & $1\farcs0$    \\ 
Sept.~30, 1999 & NTT  &  $I$  & 8$\times$300   & 1.16 & $0\farcs9$    \\ 
\end{deluxetable}
%------------------------------------------------|

%--------------------------------------
%\subsection{Stellar photometry}

The processing of the images was carried out within both the {\sc iraf} and 
{\sc midas} environments. 
After bias subtraction and flat-fielding (using twilight sky flats)
all images were registered, and sum images were produced for each
telescope by co-adding all images taken with the same filter.
The programs \daophot\ /\allstar\ (Stetson \cite{stetson87}) were run
on the sum images to obtain stellar photometry by point spread
function ({\psf}) fitting. 
%
% Care was taken to build the \psf\ in an
% iterative way using isolated stars, following the precepts of Stetson
% (\cite{stetson87}). The best results were obtained using a Gaussian
% \psf\ with a quadratic dependence on the $x$, $y$ coordinates relative
% to the frame center.
% %The final photometry catalogs in each filter were obtained by 
% %running \allstar\ twice on the sum images. 
%
For the NTT data, we also used the \allframe\ program (Stetson
\cite{stet94}), which combines \psf\ photometry carried out on the
individual images. 
% Since the photometry obtained with \allframe\
% turned out to be slightly deeper than that derived using \allstar, the
% \allframe\ measurements on the NTT images were adopted as our final
% catalog. 
The 2.2m data set was mainly employed for calibration.

%--------------------------------------
%\subsection{Calibration}
%\label{s_calib}

%------------------------------------------------------------>
%\tablecaption{Photometric zero points \label{t_zpshifts }}
%\begin{deluxetable}{lccccc}
%\tablehead{
%\colhead{Image} &
%\colhead{$X$} &
%\colhead{$t_{\rm exp}$} &
%\colhead{$\Delta m$} &
%\colhead{$\sigma_m$} &
%\colhead{$C_\lambda$} }
%%  imm sfile   airm  tesp  mediana  sigmanotte  zerop  dmedn
%%
%\startdata
% b1 & 1.024 & 900 & 0.016 & 0.005  & 7.129  \\ 
%% b2 & 1.021 & 900 & 0.008 & 0.002 & 7.138  \\ 
% v1 & 1.053 & 600 & 0.018  &0.005 & 6.789  \\ 
% v2 & 1.068 & 600 & 0.018 & 0.002 & 6.799   \\ 
% i1 & 1.023 & 900 & 0.017  &0.002 & 7.319  \\ 
% i2 & 1.031 & 900 & 0.023 & 0.004 & 7.313   \\ 
%\end{deluxetable}
%%-----------------------------------------|

Calibration of the 2.2m data was based on observations of standard
star fields from Landolt (\cite{landolt92b}). The calibration
procedure and color transformations are identical to those 
described in Held \etal (\cite{held99}). 
%
% In brief, the instrumental magnitudes of the standard
% stars, $m_{\rm ap}$, measured in circular apertures of radius $R =
% 6\farcs6$, were normalized:
% %--------------------------------------->
% \begin{equation}
% \label{e_norm}
% m' = m_{\rm ap} + 2.5\, \log (t_{\rm exp} + \Delta\,t) - k_{\lambda}\,X
% \end{equation}
% %-----------------------------------------|
% where $\Delta\,t$ is the shutter delay and $X$ is the airmass. The
% adopted mean extinction coefficients for La Silla are $k_B = 0.235$,
% $k_V = 0.135$, and $k_I = 0.048$.  A comparison of the normalized
% instrumental magnitudes with Landolt's (\cite{landolt92b})
% photoelectric photometry yields the calibration relations. For the
% 2.2m standard star observations, the rms scatter around the linear
% fits was 0.007, 0.008, and 0.008~mag in $B$, $V$, and $I$, respectively.
% %
% Instrumental \psf\ magnitudes of stars in the SagDIG field were
% converted to normalized aperture magnitudes using circular aperture
% photometry with increasing radius of bright isolated stars in the
% individual images (see Saviane et al. \cite{saviane96}).  The
% photometric zero points (which include exposure time, airmass and
% aperture correction terms) were stable, with an agreement better than
% 0.01 mag in all bands. Calibrated $BVI$ magnitudes were obtained by
% applying the relations derived from the standard stars. 
%
The total zero point errors, including the uncertainties in the
calibrations, are 0.012, 0.013, and 0.013~mag in $B$, $V$, and $I$,
respectively.

The NTT data were independently calibrated 
using the relations: 
%---
\begin{eqnarray}
B &=& b' - 0.126 \, (B-V) + 24.890 \\
V &=& v' + 0.017 \, (B-V) + 25.113 \\
V &=& v' + 0.017 \, (V-I) + 25.113 \\
I &=& i' - 0.021 \, (V-I) + 24.411
\end{eqnarray}
%---
where the $b'$, $v'$, and $i'$ magnitudes are the instrumental
magnitudes for the standard stars, measured in apertures with radius
6\farcs9, normalized to zero airmass and 1 s exposure.  The
instrumental \psf\ magnitudes of stars in the SagDIG field were
converted to normalized aperture magnitudes using a growth curve
analysis of bright isolated stars in the individual images (see
Saviane et al. \cite{saviane96}).
A comparison of stars in common with the 2.2m data set gives the
following median zero point differences: $\Delta B$ (2.2m--NTT) $=
-0.02$, $\Delta V = 0.01$, and $\Delta I = -0.05$.  The magnitude
scales are therefore in good agreement, especially for the $B$ and $V$
bands.  Since the 2.2m data have well established calibration zero
points, the latter were adopted as the reference photometric
scale. Small adjustments were applied to the NTT data accordingly.
%#
%# @4@ Le VERE differenza tra @2.2 & EMMI sono:
%# delta B(2.2-EMMI)= -0.02
%#             V          =  0.01
%#             I          = -0.05

%--------------------------------------
%\subsection{Artificial star tests}

% il file apart.dat contiene le stelle C

%---------------------------------------------------------------->
\tablecaption{
The photometric errors ($1\sigma$) and completeness from the
artificial star experiments \label{t_sigmasc}}
\begin{deluxetable}{lrrrrrr}
\tablehead{
\colhead{$BVI$} &
\colhead{$\sigma_B$} &
\colhead{$\sigma_V$} &
\colhead{$\sigma_I$}  &
\colhead{$C_B$} &
\colhead{$C_V$} &
\colhead{$C_I$}  
} 
% @3@ Le simulazione sono fatte su tutta l'immagine V, 
% non su un sub-image
%    in V & I sono state iniettate 64500 stars, mentre
%    in B                          18100 stars.
%   Bin_cent   errorB     errorV    errorI       Comp_B    Comp_V Comp_I
\startdata
18.8 & 0.01 & 0.01 & 0.01 & 0.998 & 0.998 & 0.994\\
19.3 & 0.01 & 0.01 & 0.01 & 0.999 & 0.997 & 0.994\\
19.8 & 0.01 & 0.01 & 0.02 & 0.997 & 0.997 & 0.990\\
20.3 & 0.01 & 0.02 & 0.02 & 0.997 & 0.995 & 0.990\\
20.8 & 0.01 & 0.02 & 0.03 & 0.992 & 0.994 & 0.985\\
21.3 & 0.02 & 0.03 & 0.05 & 0.992 & 0.989 & 0.980\\
21.8 & 0.02 & 0.04 & 0.06 & 0.994 & 0.984 & 0.970\\
22.3 & 0.03 & 0.05 & 0.08 & 0.985 & 0.975 & 0.945\\
22.8 & 0.04 & 0.07 & 0.12 & 0.984 & 0.963 & 0.880\\
23.3 & 0.06 & 0.10 & 0.15 & 0.967 & 0.920 & 0.825\\
23.8 & 0.08 & 0.12 & 0.17 & 0.956 & 0.860 & 0.669\\
24.3 & 0.10 & 0.13 & 0.19 & 0.940 & 0.670 & 0.450\\
24.8 & 0.12 & 0.17 & 0.23 & 0.760 & 0.450 & 0.278\\
25.3 & 0.18 & 0.21 & 0.26 & 0.550 & 0.255 & 0.196\\
\end{deluxetable}
%-------------------------------------------------------|

Photometric errors and the completeness were estimated from artificial
star experiments. Simulated stars were added to the sum images by
randomly distributing them around the nodes of a grid of triangles --
a procedure well suited to prevent self-crowding
(see Saviane et al. \cite{savi+00}). 
%
% The coordinates of the $V$ catalog were properly transformed in the
% $B$ and $I$ filters so that the relative positions of the stars are
% maintained.  The respective $(B-V)$ and $(V-I)$ colors were selected
% in specific intervals of the \cmd, for a better coverage of the
% regions of interest.
%
The frames were then reduced and calibrated as done for the original images. 
A comparison of the input and measured magnitudes for the retrieved stars 
gives the internal errors (rms in 0.5 mag bins) and completeness, 
which are given in Tab.~\ref{t_sigmasc}.

%--------------------------------------
%\subsection{Comparison with previous photometry}
%\label{s_previous}

Calibrated magnitudes for the stars in the SagDIG field were compared
with previous photometry to check the reliability of the zero
points. 
%Figure~\ref{f_compare} shows a comparison of $V,I$
%measurements for stars in common with C87.
% Two sets of stars were chosen, a bright sample of stars with 
% $V\le17.8$, and a fainter one $V\le19.2$.  
%The two data sets show good agreement. 
Our results agree with previous data of C87. The median differences
are $\Delta V= V_{us} - V_{C87}=0.020\pm0.024$, and
$\Delta(V-I)=0.030\pm0.023$ (the errors being standard deviations of
the residuals).
A direct comparison is not possible with the data of KAM99, since
photometric catalogs have not been published. However, the magnitudes
and colors of key features in their {\cmd}s (such as the \rgb\ tip or 
the blue plume) are in excellent agreement with those measured
in our diagrams.
The ($V-I$) colors measured by LK00 and by us show significant
disagreement, 
% (Fig.~\ref{f_conf.lee}), 
the median shift being
$\Delta(V-I)=0.220\pm0.086$ (this paper $-$ LK00). 
% We are inclined to be confident in our photometric scales since they
% are supported by the external checks presented in Fig.~\ref{f_compare}
% and in Held et al. (\cite{held99}).
The blue shift in the $(V-I)$ scale of LK00 significantly affects
their metallicity estimate. A systematic difference is also present in
($B- V$) colors, $\Delta(B-V)=0.100\pm0.065$ (us -- LK00).  The $V$
magnitude scales are more nearly consistent, yielding $\Delta V=
-0.040\pm0.069$.

%+++++++++++++++++++++++++++++++++++++++++++++++++++
\section{The  color-magnitude diagrams} 
\label{s_cmd}

%IS:
%
% dopo lo statistical cleaning, quanto meglio e' il nostro CMD rispetto
%  al passato? si puo' scriverci qualcosa?

%-------------------------------------------------->
\begin{figure}
\centering
\includegraphics[width=8.8cm]{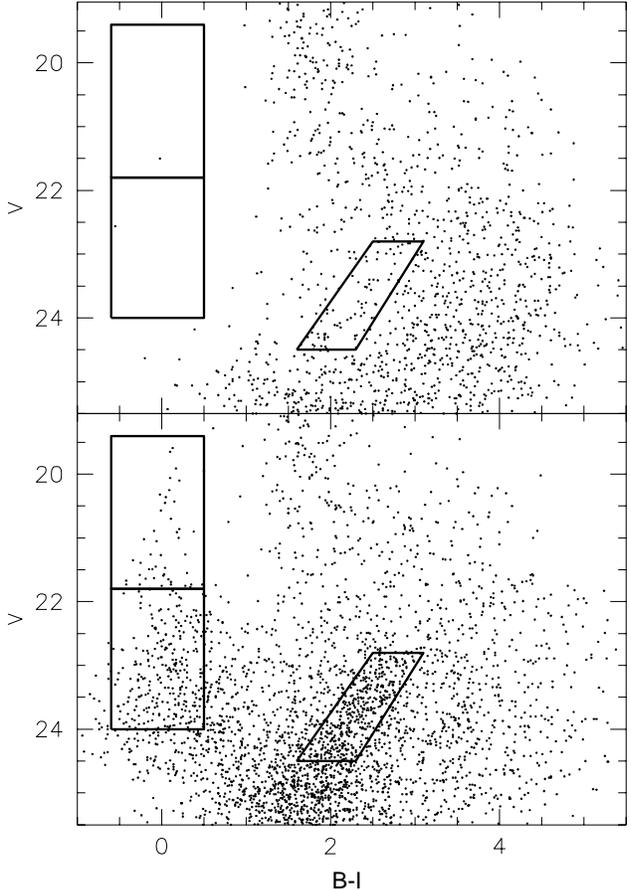}
\caption{
The $V$, $(B-I)$ color-magnitude diagram of a field centered on SagDIG  
({\em lower panel}), compared with the \cmd\ in the control region ({\em 
upper panel}). Both the inner and outer region have an area 
$5\farcm5 \times 2\farcm7$. Blue and red stars belonging to the galaxy, in 
excess over the foreground contamination, are clearly visible in the diagram 
of the inner region. The rectangles delimit two samples of young stars in two 
different age intervals, while the slanted box outlines a 
high-probability sample of  red giant stars.
}
\label{f_2contam}
\end{figure}
%-------------------------------------------------|

%------------------------------------------->
\begin{figure}
\centering
\includegraphics[width=8.8cm]{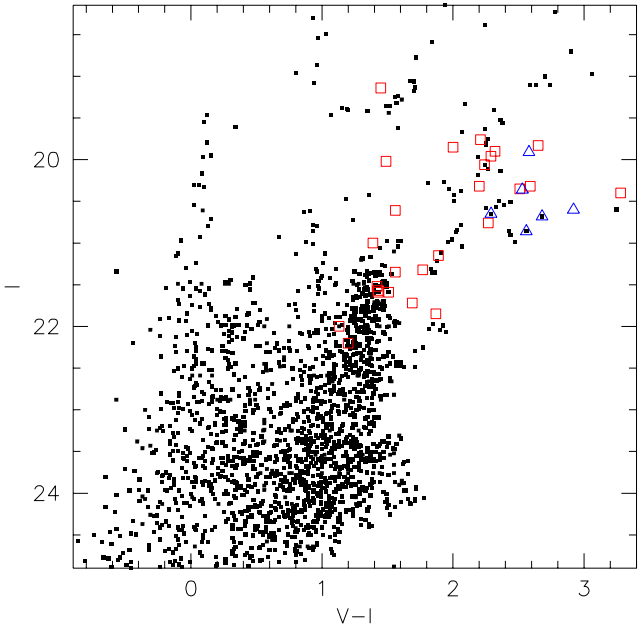}
\caption{
The $I$, $(V-I)$ color-magnitude diagram of SagDIG after applying  
statistical decontamination.
% As in Fig.~\ref{f_1pulbi}, except for the $I$, $(V-I)$ diagram. 
We have superimposed the C star candidates found by Cook
(\cite{cook87}; {\em open squares}) and those identified by Demers \&
Battinelli (\cite{deme+batt01}; {\em open triangles}).}
\label{f_1pulvi}
\end{figure}
%-------------------------------------------|

The color-magnitude diagram of the inner part of our image, comprising
most of the Sagittarius \di\ galaxy, is shown in Fig.~\ref{f_2contam}
(lower panel). The inner region was defined as 
a $5\farcm5 \times 2\farcm7$
rectangle around SagDIG, with the long side in the EW direction,
% $250 \leq x \leq 1450$ pixels and $500 \leq y \leq 1000$ pixels).
and contains both SagDIG's and foreground stars.
The \cmd\  of an outer region of equal area (upper panel) illustrates
the considerable foreground contribution of stars in the Galactic
bulge in the direction of SagDIG. Despite the field contamination, the
blue plume of young stars and the red giant branch of SagDIG are
evident in the diagram of the inner area. Three \cmd\ regions used to
select samples of stars of different ages are outlined in
Fig.~\ref{f_2contam} -- these will be utilized in
Sect.~\ref{s_stelpop} to study the stellar population gradients in the
galaxy.

A  statistical  foreground subtraction procedure  was  applied  to the
analysis of the {\cmd}s of SagDIG. All stars in the \cmd\ of the inner
region having a counterpart  in the diagram of   the outer field  were
statistically   identified   as interlopers  and     culled out.   The
coincidence was tested in  \cmd\ cells $0.3 \times  0.3$ mag  in color
and magnitude,  respectively.  An example of  the  cleaned {{CMD}} of~
SagDIG is shown in Fig.~\ref{f_1pulvi} (see also Fig.~\ref{f_isocb}).

%An example of the cleaned {{CMD}} of        
%SagDIG is shown in Fig.~\ref{f_1pulvi} (see also Fig.~\ref{f_isocb}).

All evolutionary features are more clearly seen in the decontaminated
diagrams than in the raw {\cmd}s.  The blue plume of young stars
is well delineated for more than 4 mag,
% especially those making use of the $B$ photometry (Fig.~\ref{f_1pulbi}).
% %-- and Fig.~\ref{f_1pulbv}).
% % Blue supergiant (\abbrev{BSG}) stars probably populate the region at
% % $B-V\sim0$, extending up to $V\sim19.5$ (see also KAM99).
% The presence of blue He-burning supergiant stars possibly accounts for
% the color width of the blue plume, $\sim 0.4$~mag at $V=22$,
% %--- ($B-V=0$),
% which is larger than the photometric scatter (Tab.~\ref{t_sigmasc}).
%
and the red giant branch shows a well-defined cutoff and extends
over a $\sim3$ mag interval in both $V$ and $I$. Several stars
populate the Hertzsprung gap. Although many are likely to be
He-burning stars in SagDIG, a fraction of them are possible artifacts
caused by the significant crowding in ground-based data.

Two features are noted in all of our statistically decontaminated
diagrams, so that they are likely to be real. The first is a group of
probable red supergiants located $\sim 2$ mag above the \rgb\ tip
% (see Fig.~\ref{f_1pulbi}).  
(Fig.~\ref{f_1pulvi}).  
The second feature is a tail of very red
stars with $V\sim22$, $V-I \gtrsim 1.5$, strongly resembling the
extended tail of intermediate-age asymptotic giant branch
(\abbrev{\agb}) stars found in many dwarf galaxies. This feature
suggests the presence of a significant intermediate-age population.

This conclusion is supported by the remarkable coincidence in
magnitude and color with the candidate C stars found by Cook
(\cite{cook87}) and more recently by Demers \& Battinelli
(\cite{deme+batt01}) (see Fig.~\ref{f_1pulvi}).
%
% The author identified a sample of 26 carbon stars using a
% suitable choice of intermediate band colors. 
%
The C star candidates were classified by Cook (\cite{cook87}) in two
groups: a luminous 1 Gyr old population (reaching $\sim2$ mag in $I$
above the \rgb\ tip) and a bluer, less luminous population about 10
Gyr old.
% All C stars in the C87 sample have been identified by us, and the new
% photometric data are given in Tab.~\ref{t_cstars}.
%
%IS:
%a proposito delle C stars variabili. Si dice "commonly found" che sono
%appunto variabili. Quanto sono le tipiche ampiezze? Sono comparabili con
%quelle che troviamo (suppongo di si'!)?
%
%Poiche' la I e' stata misurata quando la stella era in una fase diversa
%da quando la V e' stata misurata, il risultante V-I potrebbe avere
%scatter addizionale per questo effetto. Bisogna dunque dire che i
%periodi di queste AGB sono >> 1 giorno, ovvero piu' lunghi
%dell'intervallo in cui abbiamo fatto le osservazioni. E quindi si puo'
%considerare che le due mag sono state misurate piu' o meno nella stessa
%fase della pulsazione.
%
% Agisco di conseguenza e scrivo qualcosa qui sotto
%@@
% A comparison of C star colors and luminosities in this study and in 
% Cook (\cite{cook87}) is presented in Fig.~\ref{f_1pulvi}.
%-- in Fig.~\ref{f_cstars}.
The comparison between old and new measurements reveals an intrinsic
variability of these stars.
% , which is common for luminous intermediate-age \agb\ stars.  
The cross-identification of C stars is
firm, being based on positional coincidence within $\pm1$ pixel
($0\farcs3$). The differences between Cook's measurements and ours for
candidate C stars is, after applying a 2$\sigma$ clipping, $\Delta
V$(this paper-Cook)$=0.23$ mag with a standard deviation 0.21
mag. This scatter is significantly larger than instrumental errors
determined from artificial star experiments (see
Tab.~\ref{t_sigmasc}). Also, it is larger than the standard error of
the residuals for the general sample of stars in the same \cmd\
region, for which we measured a mean difference 0.02 mag (us$-$Cook),
with a standard deviation $\sigma_V=0.09$.
\section{Basic properties of SagDIG}
\label{s_basic}

In view of the direct dependence of the inferred galaxy distance and
metallicity on the adopted $E_{B-V}$, before discussing the basic
properties of SagDIG we must consider in some detail the reddening of
this galaxy.

\subsection{Foreground and internal extinction}
\label{s_conf_gr8}

%-- we assume 
Cook (\cite{cook87}) estimated a relatively low reddening in the direction of 
SagDIG ($E_{B-V}=0.07$) from the color distribution of the Galactic 
foreground stars. This result agrees with the prediction from the \hi\ maps 
of Burstein \& Heiles (\cite{burstein84}), $E_{B-V}=0.08\pm0.03$. The 
infrared dust emission maps of Schlegel \etal (\cite{dust98}) suggest a 
slightly higher reddening ($E_{B-V}=0.12\pm0.01$). 
%---RED STARS:
Lee \& Kim (\cite{lee00}) used their $(B-V)$, $(V-I)$ two-color
diagram to infer a very low reddening for the red giant stars in
SagDIG. Reddening values estimated using these colors, however, are
poorly constrained because the reddening vector is nearly parallel to
the stellar sequence.  
Following Cook (\cite{cook87}), we have evaluated the extinction in
the direction of SagDIG by comparing the $(V-I)$ color distribution of
the foreground Galactic stars in our database, brighter than V=22 and
redder than $B-I=0.5$, with the colors of stars in the South Polar Cap
(Caldwell \& Schecter \cite{cald+sche96}). The best fit of the two
color distributions (particularly of the blue cutoff at $V-I \sim
0.65$ related to the main sequence turnoff of Galactic disk stars) was
obtained by assuming a foreground reddening $E(B-V) = 0.07$, with an
uncertainty of about $\pm 0.02$ mag.  
A similar value, $E(B-V) \approx 0.05$, was obtained by Demers \&
Battinelli (\cite{deme+batt01}) from the $(R-I)$ distribution of
foreground stars.
%
% Our value is slightly lower than the reddening $E(B-V) = 0.12$ derived
% using the Schlegel's et al. (\cite{dust98}) infrared maps.  A value as
% high as $E(B-V) = 0.2$ is definitely ruled out.

%--BLUE STARS:
Evidence for a higher reddening has been found in the
star forming regions. In their spectroscopic study of \hii\ regions in
SagDIG, 
% IS: c _non_ e' il rapporto Hbeta/Halpha
Skillman et al. (\cite{stm89}) obtained $c(\rm H \beta) = 0.33$. 
%--- Orig
%Skillman et al. (\cite{stm89}) measured the ratio
%$c(\mbox{H}\beta) = \mbox{H}\beta/\mbox{H}\alpha$, obtaining $c(\rm H
%\beta) = 0.33$. 
%--
This value was converted into $E_{B-V}=0.22$ using the
relation $c(\mbox{H}\beta)=1.47\times E_{B-V}$ (Seaton \cite{seaton}).
Recent measurements of the Balmer decrement in the largest \hii\
region by Saviane et al. (\cite{savi+01}) confirms this high
reddening, yielding $E(B-V)=0.19 \pm 0.04$. 
%IS:
%E' Giusto Chiamare L'ISM Ionizzato "Warm"?
While this result strictly
pertains to the warm \ism, it is likely to apply also to the young
stars in the blue plume, physically associated to the interstellar gas.

The presence of a differential extinction between young and old stars
is well known in starburst galaxies (e.g., Calzetti et
al. \cite{calz+94}). Recently formed 
stars are embedded in regions of high dust content, while old stars
are distributed in more extended morphological structures
(halos or disks) (e.g., Zijlstra \& Minniti \cite{zijl98}).
A differential reddening has been suggested in some
other {\di}s, for instance IC~10 (Sakai et al. \cite{sakai+99}). 
%
% In that case, a difference $\Delta E_{B-V} \sim 0.3$ for the reddening
% of the young and old population was invoked to reconcile the distances
% derived from the Cepheid variables and the \rgb\ tip. 
%
This might well be the case also for SagDIG, where the young stars are
concentrated near the \hi\ density peaks (see Sect.~\ref{s_spatial}),
while the older stars show an extended distribution.  

% It remains to be explained why this phenomenon apparently does not
% affect the stellar population of \grotto. The different behavior might
% be related to the fact that the densest gaseous regions of \grotto\
% are found outside the main optical body of the galaxy, while in IC~10
% and SagDIG the star forming regions are found near the densest \hi\
% clumps (Wilcots \& Miller \cite{wilcots98}; Carignan et
% al. \cite{carignan90}).

\subsection{Distance and metallicity}
\label{s_rgb}

%%----------------------------------------------->
\begin{figure}
\centering
\includegraphics[width=8.8cm]{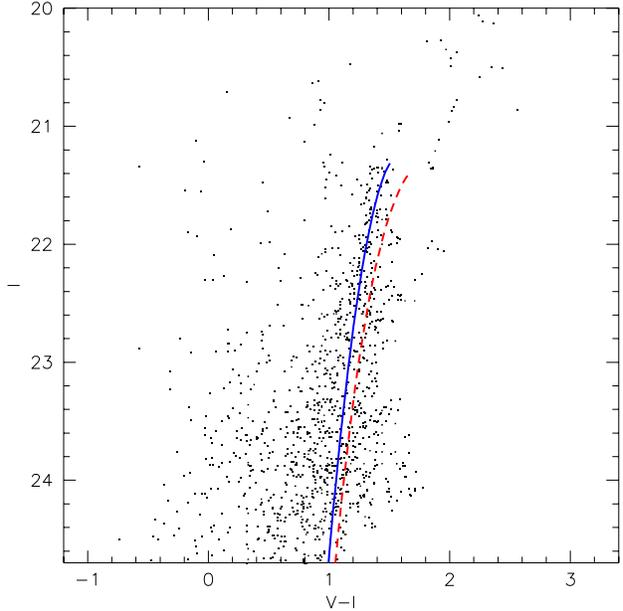}
\caption{
The $I$, $(V-I)$ color-magnitude diagram of a sample of SagDIG stars with 
distance from the galaxy center $r > 0\farcm9$. The red giant stars in
SagDIG are nearly as metal-poor as those in 
the Galactic globular cluster M\,15 ([Fe/H]=$-2.17$, {\em 
solid line}), and more metal-poor than stars in 
M\,2 ([Fe/H]=$-1.58$, {\em dashed line}). We use a distance 
modulus $(m-M)=25.14$ and a reddening $E_{B-V}=0.07$ for SagDIG.
}
\label{f_rgbsamp}
\end{figure}
%--------------------------------------------|

% Figure~\ref{f_fl_vi} shows the $I$-band magnitude distribution
% (hereafter \abbrev{LF}) of the red giant stars in SagDIG.  
%
The $I$-band magnitude distribution
(hereafter \abbrev{LF}) of the red giant stars in SagDIG 
was obtained from a statistically decontaminated sample of stars 
selected outside a radius $r= 0\farcm9$ from the galaxy center.
%-- galaxy center at x=1045   y=950   pixel=0.27 
This choice excluded most star-forming regions, yielding a clean
sub-sample of red giants in the ``halo'' of SagDIG
(Fig.~\ref{f_rgbsamp}). The LF was derived by simply counting stars in
the color range $0.8 < V-I < 1.8$.

%%%--------------------------------------
%\subsection{Distance} 
%\label{s_dist} 

The clear \rgb\ cutoff seen in the LF was identified as the tip of the
red giant branch, and employed to derive the distance to SagDIG
according to the iterative methods of Mould \& Kristian
(\cite{mould86}) and Lee et al. (\cite{lee93}) (see Held et
al. \cite{held99}).
The \rgb\ tip location was measured by filtering the luminosity
function with a Sobel kernel (see, e.g., Madore \& Freedman
\cite{mado+free95} and references therein). This method yielded
$I_{\rm tip}= 21.30 \pm 0.15$, where the uncertainty includes the
internal error (assumed to be half the bin size) and the calibration
errors.  As a check, we also measured the \rgb\ cutoff by fitting an
error-convolved step function to the magnitude distribution, obtaining
the same result.  Our estimate of the tip level is therefore slightly
brighter than that obtained by KAM99 ($I_{\rm tip} = 21.38$), although
the difference is within the quoted errors.

% IS: 
%Si dice che MI(rgbt)=-4.0, cosa vera fino ad un po' di tempo fa. Pero'
%un po' di revisioni teoriche ed osservative, farebbero ora pensare a un
%-4.2 (cf. paper di Harris et al. su Nature, distanza di Virgo
%cluster). 
%D'altra parte la calibrazione di Bellazzini et al. 2001 ritorna al
%vecchio valore... insomma, ci metto qualcosa
%
The initial guess of the distance was based on the assumption that
$M_I^{\rm RGBT}=-4.0$, which is an excellent approximation for
metal-poor systems, where the $I$ luminosity of the \rgb\ tip shows
only a modest dependence on the galaxy metallicity (see also Salaris
\& Cassisi \cite{sc98}; Bellazzini et al. \cite{bfp01}).
%
% This property allows refining the estimates of distance and metallicity
% in an iterative way.
% The adopted luminosity of the RGB tip is based on the theoretical HB
% luminosity predicted by the Lee et al. (\cite{ldz90}) models. More
% recent theoretical models (Salaris \& Cassisi \cite{sc98}) and GGC
% distance determinations based on the local subdwarfs measured by
% Hipparcos (Gratton et al. \cite{gratton_etal97}), would predict a
% brighter tip luminosity $M_I^{\rm RGBT}\simeq -4.2$. However, a
% determination closer to the old distance scale has been determined by
% yet another recent investigation (Bellazzini et al. \cite{bfp01}), so
% for the sake of comparison with currently available galaxy distances, we
% will keep the ``long'' distance scale in the present analysis.
%
Our distance estimate assumes a low reddening
($E_{B-V}=0.07 \pm 0.05$, $E_{V-I}=0.09\pm 0.06$, $A_{I}=0.13 \pm 0.10$) 
for the \rgb\ population, and a mean metallicity [Fe/H]=$-2.08 \pm
0.20$ dex (see below). With these assumptions, we obtained a corrected
distance modulus $(m-M)_0 = 25.14 \pm 0.18$ ($1.07 \pm 0.09$
Mpc), a value confirming previous estimates. The distance error
includes all uncertainties on the \rgb\ tip magnitude, reddening and
metallicity through standard error propagation.

%%--------------------------------------
%\subsection{Metallicity}
%\label{s_feh} 

The mean [Fe/H] of the ``old'' stellar population in SagDIG was
estimated as in Saviane et al.  (\cite{saviane96}), by comparing the
colors of the red giants with the \rgb\ fiducial sequences of template
Galactic globular clusters (GGC).  The adopted reddening values are
those specified above.
% The median \rgb\ color at $M_I=-3.5$ was measured from stars in the
% intervals $I=21.8 \pm 0.5$, $1.0 \leq (V-I) \leq 2.0$.  We obtained
% $(V-I)_{-3.5} = 1.41 \pm 0.02$.
%-- actually \pm 0.015
% Using $E_{V-I}=0.09$, this mean color corresponds to a
% reddening-corrected color $(V-I)_0=1.32 \pm 0.07$.  
%Using the interpolation formula of Lee et al. (\cite{lee93}), this 
%corrected color implies [Fe/H]$-1.77 \pm XXX$ dex. 
%-- [Fe/H]$= -12.65 + 12.6\,(V-I)_{-3.5} - 3.3\,(V-I)_{-3.5}^2$, 
%
The average color differences between the \rgb\ stars in SagDIG and
the GGC fiducials, measured in the interval $21.3 < I < 22.3$,
were measured using the cluster ridge lines and
metallicities of Da Costa \& Armandroff (\cite{da90}).  
%-- , are plotted in Fig.~\ref{f_cmd90}.  
%
A quadratic fit to the relation between mean color shifts and GGC
metallicities yields [Fe/H]=$-2.08$ for SagDIG.  The uncertainty,
computed from the error on the color difference, is about 0.20 dex.
% %
% We note that the presence of a young stellar population on the blue
% side of the \rgb\ could in principle bias the measured \rgb\ color.
% In practice, the fraction of young helium-burning stars in this sample
% with $r > 0\farcm9$ does not appear sufficient to affect the {\it
% median} \rgb\ color.

%-- metodo di IVO:
% In addition, we measured the metallicity from the slope of the \rgb\
% following the method of Saviane et al. (\cite{savi+00b}).  We found
% again [Fe/H]$= -2.08\pm0.15$ using the color distribution of 90 stars
% in the interval $-4 < M_I < -1$, for an assumed reddening
% $E_{V-I}=0.09$.

The metallicity obtained for SagDIG is among the lowest for dwarf
galaxies, yet less extreme than found by previous authors.  Had we
adopted a reddening $E_{B- V}=0.12$ (i.e. $E_{V-I}=0.15$) as in KAM99,
a mean [Fe/H]$= -2.31$ would have been obtained, which indicates that
our metalliciy estimate and that of KAM99 are essentially consistent
given the different assumption for the reddening.
The much lower metallicity range inferred by LK00 is most
likely related to the photometric zero points adopted by these
authors, yielding $(V-I)$ colors which are systematically bluer than
those obtained by C87, KAM99, and this study.

% Our metallicity estimate is higher than that
% found by Karachentsev et al (1999) and by Lee \& Kim (2000). We note
% however, that Karachentsev et al. used a slightly higher reddening
% value ($E_{B-V}=0.12$) reflecting in a metallicity estimate of
% [Fe/H]$=-2.5$, whereas Lee \& Kim (2000), assuming a zero reddening,
% derived [Fe/H]$=-2.4$.  In the light of photometric comparison
% presented in Sect. \ref{s_previous}, we suggest that the Lee \& Kim
% (2000) data might suffer calibration problems, moreover, had we
% adopted a reddening of $E_{B-V}=0.12$, as in Karachentsev et al., we
% still would have obtained a ``normal'' [Fe/H] of $-2.31$.  We
% therefore conclude that our metallicity is more appropriate.  

%+++++++++++++++++++++++++++++++++++++++++++++++++++++++++++++++++++++
\section{Stellar populations}
%\section{Discussion}
\label{s_stelpop}

% IS:
%- nella recent SF ci _deve_ essere un contributo a <5 Myr. Senza
%  togliere il burst a 30 Myr, dopo di questo ci deve essere stata SF meno
%  accentuata (bursting? gasping?), la LF delle stelle giovani non lo
%  esclude. Il nr. di stelle formato e' molto piccolo, sicche' non si
%  notano le stelle post-MS

%--------------------------------------------------
\subsection{The recent star formation}
\label{s_sfh}

%------------------------------------------------->
\begin{figure}
\centering
\includegraphics[width=8.8cm]{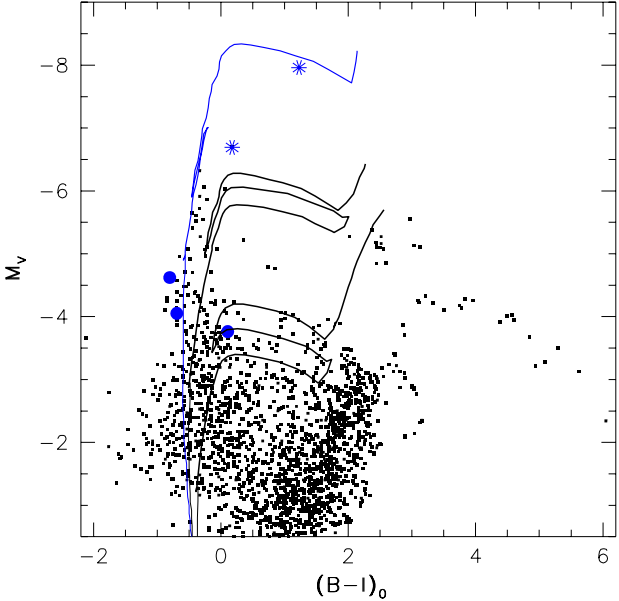}
\caption{
The color-magnitude diagram of young stars in SagDIG, with superposed
model isochrones with $Z=0.0004$ and ages $\sim$10, 30, and 100 Myr
(from Bertelli et al. \cite{bertelli94}). We adopt a reddening
$E_{B-V}=0.19$, appropriate for the star-forming regions, and an
apparent distance modulus 25.73. The {\em thin lines} represent the
core H burning phase, while the {\em thick lines} represent the faster
post-MS stages. Large filled circles represent three stars identified
in the largest \hii\ region, while the two starred symbols are objects
coincident with the two smaller \hii\ region candidates.}
\label{f_isocb}
\end{figure}
%------------------------------------------|

%XXX=VEDI ANCHE SEXTANS A PER L'ASSOCIAZIONE CON L'IDROGENO=XXX \\

Figure~\ref{f_isocb} shows a statistically cleaned color-magnitude
diagram with superimposed a set of theoretical isochrones with ages
10, 30, and 100 Myr. 
The isochrones are $Z=0.0004$ models ([Fe/H]=$-1.7$) from Bertelli's
et al. (\cite{bertelli94}).  A thick line emphasizes the He burning
part of the isochrones.  The \cmd\ was transformed into the $M_V$,
$(B-I)_0$ plane by adopting a true distance modulus 25.14 (see
previous section) and a reddening appropriate for the star forming
regions ($E_{B-V}=0.19 \pm 0.05$, $A_{V}=0.59 \pm 0.10$). The apparent
distance modulus is therefore 25.73 mag.
%---XXXXXXX--NB. 25.36 sbagliato=wrong

The general picture of star formation is quite similar to that
delineated by previous work. However, the cleaner CMD allowed by our
statistical decontamination, together with the larger photometric
baseline, provide new interesting pieces of information on the
\abbrev{SFH} of SagDIG.
First, the blue-loops of the 30 Myr isochrone match quite well the
luminosity and color of the two groups of blue and red stars near $M_V
\simeq -5.5$. This supports the reality of the red group and its
identification as a clump of red supergiants.
% These clumps appear to be well fitted by a metallicity 
% close to $Z=0.001$. 
According to the Padua models, such age implies a MS turnoff mass
$\sim$ 9 M$_{\odot }$.  
Secondly, the blue stars in the range $-5<M_{V}<-4$ appear perhaps too
blue for $He-$burning stars, as KAM99 suggested -- it is quite
possible that they are MS stars slightly older than 10 million
year. Further, we note the absence of a prominent plume of red
supergiants that should be expected to define the red side of the blue
loops. HST observations may eventually clarify this issue.
%
% IS:
% questo lo toglierei, perche' secondo me l'ultimo episodio di SF
% deve essere piu' giovane di almeno 5 Myr... vedi telefonata
%
%The lack of stars brighter than $M_V=-6$ seems to rule out
%significant star formation in the last 10 Myr.
% The blue loops of 100 Myr old stars having $Z=0.001$ appear to be too
% extended, possibly implying a lower metallicity (e.g., $Z \leq
% 0.0004$). 
% Note, however, that since we have applied a high-reddening
% correction throughout the diagram, and a differential reddening is
% likely to be present in SagDIG, the blue-shift of RGB stars in this
% figure was overestimated.
Also, the obvious shortage of stars with age between 30 and 100 Myr
suggests that the star formation rate dropped nearly 100 million years
ago, with a short burst of star formation about 30 Myr ago
interrupting a relatively quiescent period.
%- From a synthetic \cmd\ analysis, KAM99 suggested that the
%- galaxy is currently experiencing a high star forming activity epoch,
%- at a rate of 10 times greater than the average for its entire life.
%
% The stars filling the Hertzsprung gap, $\sim 0.3$~mag redder than the
% MS locus, are consistent with the expected location of helium burning
% (HeB) stars with age older than $10^8$ yr. However, ground-based data
% cannot constrain the SF behavior before 100~Myr, although we expect
% such a small galaxy to have a bursting SF along its entire life.  The
% absence of blue loop stars younger than 30 Myr indicates a sudden
% decline in the star formation rate (\sfr) nearly 30 million yr ago.
%
%%-----------------------------------------
%\subsection{\hii\ regions}
%
% IS:
%Ricordiamo che i "two smaller candidate Hii regions" potrebbero essere
%foreground objects secondo Skillman et al.

Finally, we remark the presence of a few very bright stars
spatially coincident with the candidate \hii\ regions 
investigated by Strobel et al. (\cite{strobel91}) 
(Table~\ref{t_stars}). 
Two bright objects ($M_V < -6$) were found in each of the two smaller
candidate \hii\ regions (\#1 and \#2). These objects have been
classified as probably galactic, given the absence of [\oiii] emission
(Skillman et al. \cite{stm89}). The brightest one (in region
SHK~1) is quite red, which is consistent with the identification with
a foreground dwarf star. The
color of the star coincident with region SHK~2 is consistent with
a supergiant or a star clusters unresolved in ground-based
data (see Cappellari et al. \cite{capp+99}). Three stars coincident in
space with the bona-fide \hii\ region (\#3) are luminous and hot
enough to represent possible exciting stars. 
% IS:
%
%"filled squares in Fig.12" -> sono _tutti_ filled squares, quindi sono
%large filled squares (ma un simbolo diverso forse sarebbe meglio)
They are as luminous as expected for $\sim20$ Myr main-sequence stars
($M_V \sim -4$, large circles in Fig.~\ref{f_isocb}).

\tablecaption{Properties of stars projected onto the \hii\ regions
\label{t_stars}}
\begin{deluxetable}{lrrrrcc}
\tablehead{
\colhead{region} &
\colhead{star}   &
\colhead{$x$}   &
\colhead{$y$}   &
\colhead{$V$}   &
\colhead{$B-V$} &
\colhead{$V-I$}  
}
\startdata 
% reg              x       y       V       BV      VI 
{\sc shk} 1 &  12850 & 875.9 & 790.0 & 18.01 & \phantom{$-$}0.77  & \phantom{$-$}0.89 \\ %-7.2
{\sc shk} 2 &  14120 & 859.9 & 869.8 & 19.28 & \phantom{$-$}0.14  & \phantom{$-$}0.47 \\ %-6.1
{\sc shk} 3 &  12880 & 1182  & 792.5 & 21.35 & $-$0.18 & $-$0.19 \\ %-4.0
{\sc shk} 3 &  12820 & 1166  & 788.1 & 22.21 & \phantom{$-$}0.05  & \phantom{$-$}0.49 \\%-3
{\sc shk} 3 &  13150 & 1185  & 809.6 & 21.92 & $-$0.16 & $-$0.10 \\ %-3.4
% La regione 3 e` quella studiata da IVO
\end{deluxetable}
\subsection{Spatial distributions: young stars vs. red giants}
\label{s_spatial}

% IS:
%
%a proposito delle distribuzioni spaziali. ad un certo punto si era
%notato che le stelle giovani sono distribuite lungo una specie di barra,
%con le due regioni piu' attive agli estremi, come spesso si nota nelle
%dirr e non. questo non va piu' bene? eppure basta guardare l'immagine
%a 3 colori...
 
%--------------------------------------------------->
\begin{figure*}
\includegraphics[width=11.5cm]{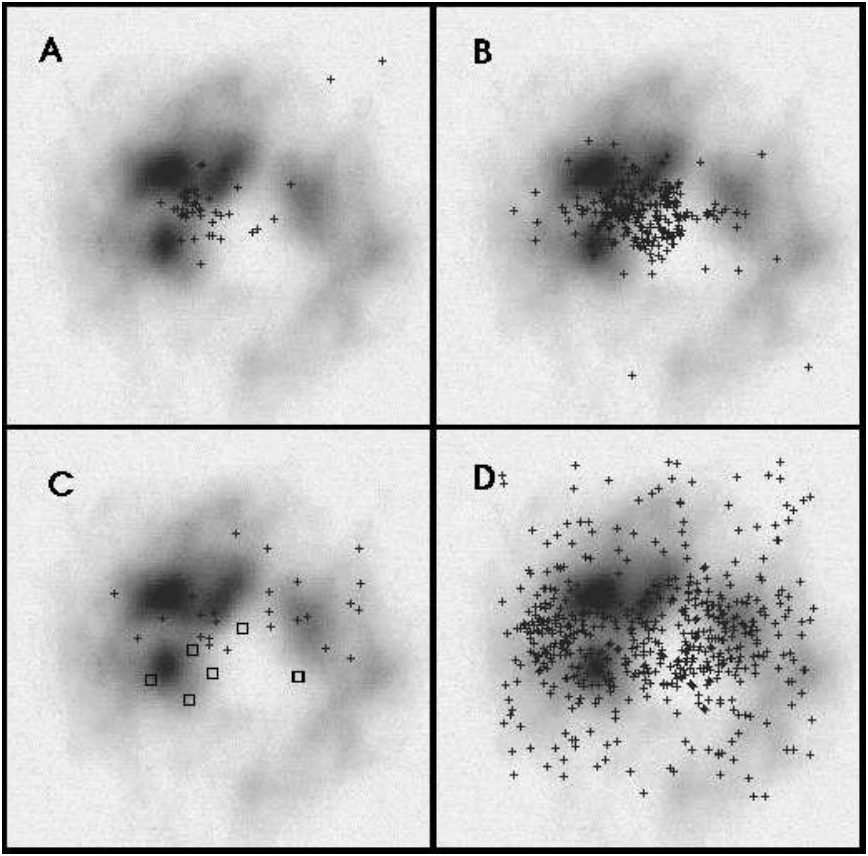}
\hfill
\parbox[b]{55mm}{
\caption{ 
The spatial distribution of young and old stars in SagDIG superimposed
on the surface density map of \hi\ from Young \& Lo (\cite{young97}).
{\bf a)} The most luminous blue stars in SagDIG, including both MS and
{\abbrev BL} stars;
{\bf b)} main sequence stars with age between 40 and 200 Myr;
{\bf c)} the carbon stars candidates from Cook (\cite{cook87}, {\it
crosses}) and Demers \& Battinelli (\cite{deme+batt01}, {\it
squares});
{\bf d)} red giant branch stars as selected in Fig.~\ref{f_2contam}.
The field size of each panel is 8.5 arcmin. N is to the top, E to the
left.  }
\label{f_both}
} %end-parbox
\end{figure*}
%--------------------------------------------|

%----------------------------------------------------->
\begin{figure}
\centering
\includegraphics[width=8.8cm]{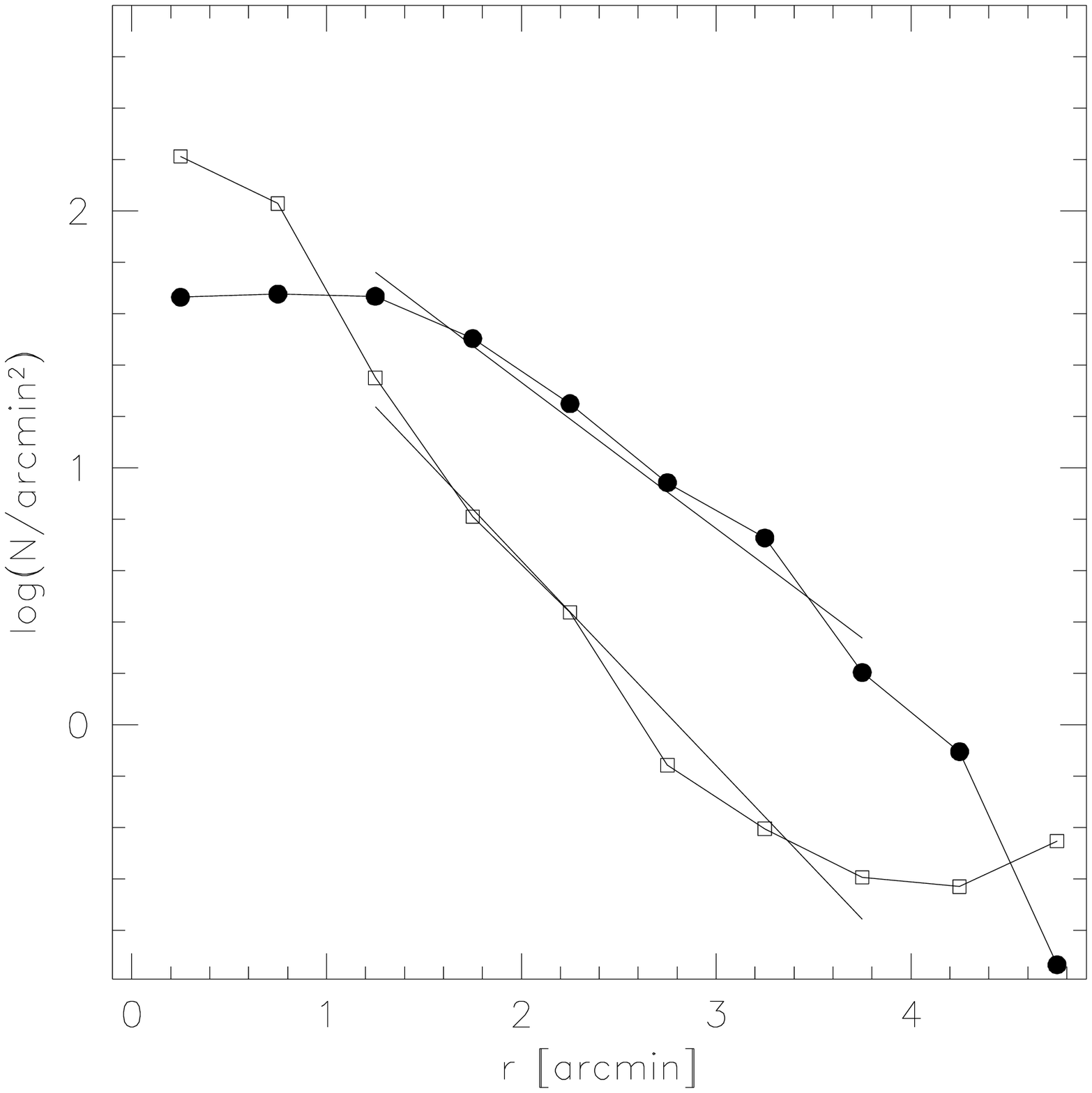}
\caption{ 
The surface density profiles of young stars ({\em open squares}) and
red giant stars ({\em filled circles}) in SagDIG. The straight lines
represent exponential fits to the density distribution of young and
old stars. Note the larger scale length of the extended halo of red
stars surrounding the inner, star-forming regions of SagDIG. }
\label{f_2boxes_profili}
\end{figure}
%-------------------------------------------|

Stars of different ages in SagDIG are distributed in a very different
way across the galaxy. Previous studies, starting with Cook
(\cite{cook87}), had already noted that blue stars are only found in
the central regions of SagDIG. In the following, we will show that the
youngest stars are embedded in a region of high \hi\ density, while
the older populations are located away from the galaxy center, forming
an extended halo or disk with a radial exponential fall-off.

Figure~\ref{f_both} shows the spatial distribution of stars in
different age intervals, superimposed onto the surface density
distribution of the neutral gas (from Young \& Lo
\cite{young97}).  The latter authors showed that the \ism\ of SagDIG
comprises a broad, warm component ($\sigma=10$ \kms) and a narrow,
cold one ($\sigma = 5$ \kms), with the warm component distributed
throughout the galaxy (out to 3 Kpc from the center) while the cold
one is concentrated into a small number of clumps of $\sim8\times10^5$
\msol.  A prominent clump is nearly coincident with the biggest \hii\
region (Strobel et al. \cite{strobel91}). 

The color and magnitude selections are those outlined in
Fig.~\ref{f_2contam}; no further statistical decontamination or
spatial selection was applied. Two boxes sharing the color
interval $-0.5<B-I<0.5$ were used to pick up MS stars approximately
between $20-80$ Myr ($21.8<V<24.0$) and younger than 20 Myr
($19.4<V<21.8$), respectively.  Both regions will also contain a
fraction of He burning stars on the blue loops, somewhat older than MS
stars at a given luminosity. For example, the blue stars in the
brighter subsample are a mixture of MS stars younger than 20 Myr and
He burning stars with age $<100$ Myr.
% 
% For example, the most luminous blue stars in Fig.~\ref{f_isocb} are
% likely to be 30-Myr old He-burning stars rather than 5-10 Myr old main
% sequence stars.
%

The distributions of the young stars in both age intervals are plotted
in Fig.~\ref{f_both}$a$ and $b$.  The location of the youngest stars
(panel $a$) indicates that the most recent star formation episode
occurred in a small region located between the two brightest \hi\
clumps along the gaseous shell, somewhat offset from the \hi\ hole.
The size of this association is $\sim 0\farcm5$ (180 pc).
%
% The hypothesis that the \hi\ clumps are associated with star formation
% (Young \& Lo \cite{young97}) finds support in these results.
% The \hii\ regions studied by Strobel et al.  (\cite{strobel91}) are
% also projected near the \hi\ peaks. 
% These results put on a firm quantitative basis the conclusions of
% Young \& Lo's (\cite{young97}) based on visual inspection of the
% Palomar Sky Survey.

The less young stars show a clumpy distribution around the
same region, with a lower central concentration and a spatial extent
of the order 2\arcmin\ in diameter ($\sim 600$ pc). If we make the
hypothesis that the young stars retain the velocity dispersion of the
cold, dense \hi\ clumps from which they probably originated, this size
appears consistent with star diffusion in $\gtrsim100$ Myr.  Indeed,
with a stellar velocity dispersion of about 5 \kms, equal to that of
the narrow \hi\ component (Young \& Lo \cite{young97}), the crossing
time of a 0.5 Kpc region is 100 Myr.  Note, however, that the
asymmetric shape of the star forming region in Fig.~\ref{f_both}$b$, as
well as the presence of several clumps in the distribution of young
stars, suggest the occurrence of stochastic star formation associated
with the \hi\ shell.
% (see, e.g, REFSXXX).
%
Probably what we are seeing is the result of a combination of the two
mechanisms. 
%
% Star formation appears to have migrated around the center of gravity
% of the \hi\ cloud toward the outer regions following the expansion of
% the shell.
%
We finally note that some blue stars are found in ``tails'', perhaps a
hint of spiral structure, quite far from the main star-forming site.
This may be the signature of star formation on the rim of the shell.
%- One should also bear in mind the possibility that ``voids'' are the
%- effect of patchy absorption clouds.

%--xxxevh:
%  1 km/s => 3.15576e13 km/Myr => 1.02 pc/Myr (1 pc = 3.09e13 km)
%  Then: 5 km/s => 5.1 pc/Myr => 0.51 kpc/100 Myr. 
%  nb. 1' = 310.26 pc, 1" => 5.171 pc
% le dimensioni delle regioni sono misurate a occhio: 35" youngyoung,
%  115" less young, 4.8'x2.5' red giants. 
%--DISCUTERE:
%   ASSOCIAZIONE DELLA STAR FORMATION COL BORDO DELLA
%   SHELL ; CONFRONTARE CON ALTRE GALASSIE;
%  IN PARTICOLARE: COOL VS WARM PHASE IN YL97;
%  TIMESCALE DELL'ESPANSIONE DELLA SHELL E DELLA 
%  DIFFUSIONE DELLE STELLE, VEDI ALTRE GALASSIE;
%  DIMENSIONI DELLE ASSOCIAZIONI. 

%---------------------------------------
% --- RGB AND CARBON STARS: -----------
In contrast to the distribution of the young stars, the red giants in
SagDIG are uniformly distributed across the galaxy
(Fig.~\ref{f_both}$d$), with their mean position approximately
coincident with that of the blue stars. Interestingly, we note a
degree of coincidence with the ridges and arms in the distribution
of neutral gas, which appears more than fortuitous -- e.g., note the
arm-like filaments of stars tracing the western rim of the \hi\
shell. These red stars need not be very old: they may well be
relatively young red giants (yet older then 1 Gyr), 
as it is probably the case in Leo~A
(Tolstoy et al. \cite{tolstoy98}). 
The intermediate-age upper-\agb\ population seems to share the
extended distribution of the red giants, as it can be seen from the
distribution of C stars in the samples of Cook (\cite{cook87}) and
Demers \& Battinelli (\cite{deme+batt01}) (Fig.~\ref{f_both}$c$).

%--XXXEVH: 
%   1) CI SONO STELLE C NEL BULGE ? VEDI NG.
%  2) DISTRIBUZIONE DI LUMINOSITA', VEDI AD ES. SMC

%------------------------------------  PROFILI DI DENSITA' --------
%
%==
%quando si parla delle scale lengths, sarebbe carino confrontare con
%valori tipici per dIrr e BCDs, e' noto che sono diverse, per cui si
%potrebbe vedere a quale categoria appartiene SagDIG. Mi sono stampato
%un po' di papers, ma non ho ora il tempo di leggerli e scriverci
%qualcosa. O lo lasciamo per la versione riveduta, o se vuoi scrivere
%qualcosa, le refs che ho trovato sono: 
%
% Marlowe et al. 1999, ApJ, 522, 183
% Eon-Chang et al. 1998, ApJ, 505, 199
% Papaderos et al. 1996, A&A, 314, 59
%

Figure~\ref{f_2boxes_profili} shows the different surface density
profiles of blue stars (younger than about 80 Myr; $19.4<V<24.0$) and
red giants. The profiles are plotted against the distance from the
galaxy center, defined as the mean $x$, $y$ of red giants. Both
profiles are nearly exponential as expected in dwarf galaxies. It is
apparent from this figure that the radial scale length of the \rgb\
stars is larger than for the blue stars. A linear fit in the radial
range $1\farcm2 < r < 3\farcm8$ (0.4--1.2 Kpc) provides a quantitative
estimation of the different exponential scale lengths of young and
``old'' stars. The scale lengths are 46\arcsec\ and 33\arcsec\ (0.24
and 0.17 Kpc) for the red giants and blue stars, respectively.
% IS:
%
%si dice che KAM99 e LK00 hanno scale lenghts diverse, ma non si citano,
%e non si dice perche' sono dominate dalle young stars
The
global scale lengths measured by KAM99 and LK00 are clearly dominated
by the young stars. The central flattening of the \rgb\
star profile is quite uncertain in the innermost 0.3 Kpc, because of
incompleteness in the star counts caused by the extreme crowding. In
particular, the red stars are likely to be masked by the more luminous
young star complexes. 
%- The profiles are also somewhat ill-defined in
%- the outer regions because of the poor statistics.
%
Overall, there are several hints that SagDIG may be a disk galaxy seen
nearly {face-on}, owing its irregular shape to the optical
signatures of recent star formation. The rotation required to support
such a low-mass galaxy is very small, and could easily be masked by
the turbulent gas motions (see Young \& Lo \cite{young97}).

%--------------------------------------------------------------------
\subsection{The metallicity-luminosity relation}

\realfiguremid{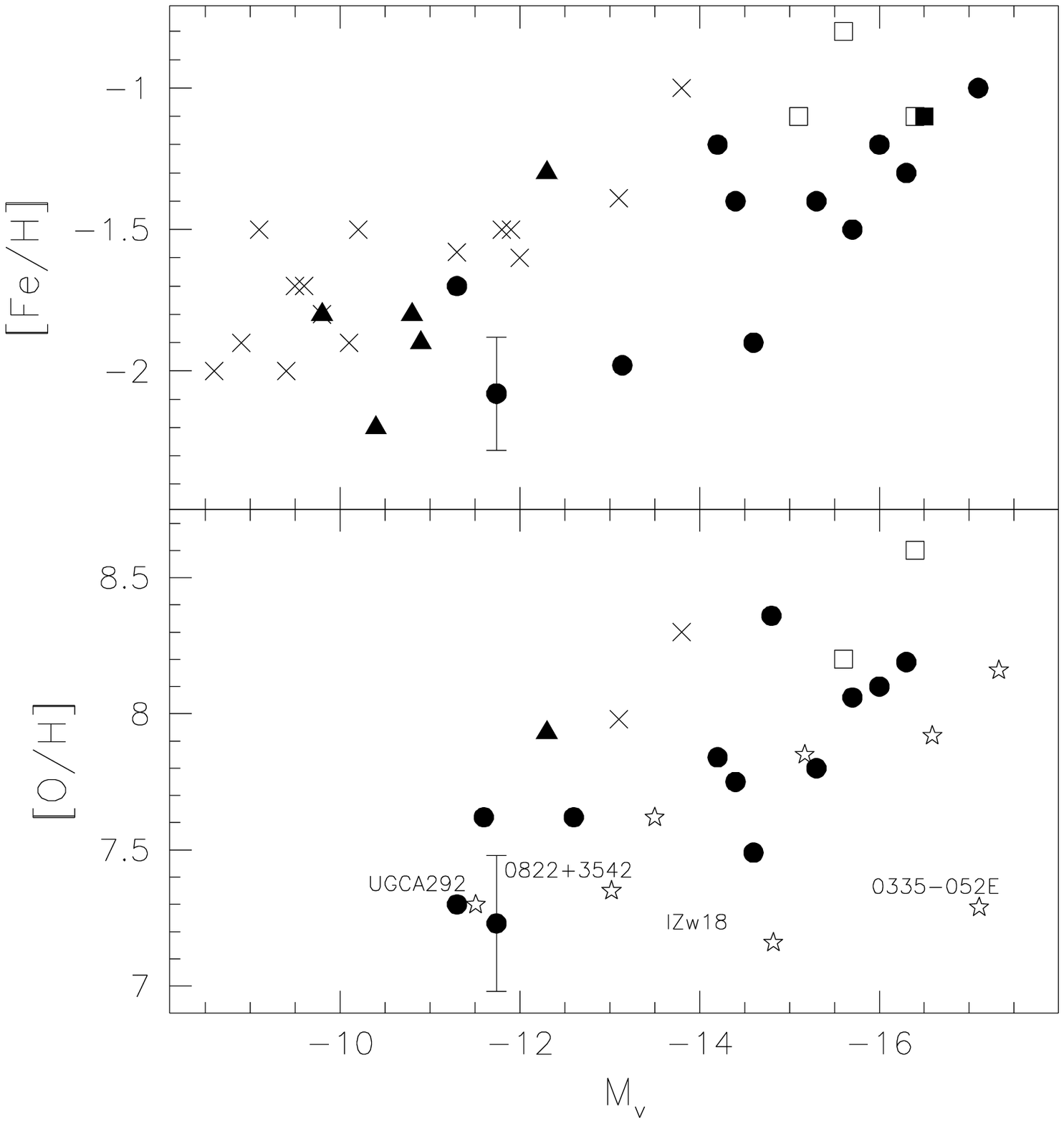}{Metallicities of stars and the \ism\ of
dwarf galaxies in the Local Group and beyond, plotted against their
$V$ absolute magnitude. References to data are given in the text.
{\em Upper panel:}\ {\em crosses} refer to dSph galaxies, {\em open
squares} are the dE companions to M\,31, the {\em filled square} is
the data point for the low-luminosity elliptical M\,32, {\em filled
circles} are dwarf irregulars, and {\em triangles} represent
``transition'' galaxies (dSph/dI) (see Mateo \cite{mateo98}).
SagDIG is represented by the data point with error bars.  
{\em Lower panel:} The [O/H] abundances of the gaseous component, measured
from nebular emission lines. The symbols are the same as for the
stellar metallicities. {\em Stars} represent dwarf irregulars or blue
compact dwarfs outside the Local Group.}{f_mateo}
% %----------------------------------------------->
% \begin{figure}
% \centering
% \includegraphics[width=8.8cm]{MS1759f7.eps}
% \caption{
% Metallicities of stars and the \ism\ of dwarf galaxies in the
% Local Group and beyond, plotted against their $V$ absolute
% magnitude. References to data are given in the text.  {\em Upper
% panel:}\ {\em crosses} refer to dSph and dE galaxies, {\em filled
% circles} are dwarf irregulars, and {\em triangles} represent
% ``transition'' galaxies (dSph/dI) (see Mateo \cite{mateo98}). The {\em
% square} is the data point for the compact low-luminosity elliptical
% M\,32. SagDIG is represented by the data point with error bars. 
% {\em Lower panel:} The [O/H] abundances of the gaseous
% component, measured from nebular emission lines. The symbols are the
% same as for the stellar metallicities. {\em Stars} represent dwarf
% irregulars or blue compact dwarfs outside the Local Group.}
% \label{f_mateo}
% \end{figure}
% %---------------------------------------------|

The metallicity estimated in Sect.~\ref{s_rgb} for the red giant stars
in SagDIG ([Fe/H]$\simeq-2.1$) turns out to be in good agreement with
the metal content estimated for the ISM.  A new determination of the
oxygen abundance in the \hii\ regions of SagDIG is presented in a
companion paper (Saviane et al. \cite{savi+01}).  The new estimate,
$12+\mbox{log(O/H)}=7.23\pm 0.20$, revises downwards the [O/H]
determination of Skillman et al. (\cite{stm89}), yielding (under the
same physical assumptions) a value more metal-poor by about 0.2 dex.
We have transformed this result to [Fe/H]=$-2.07 \pm 0.20$ by adopting
$12+\mbox{log(O/H)}_{\odot}=8.93$ for the solar O abundance, and using
the mean difference between [Fe/H] and [O/H] given by Mateo
(\cite{mateo98}).
Under these
assumptions, the composition of the warm gas in the \hii\ regions does
not appear to be significantly metal enriched with respect to the red
giant stars.

These new estimates of the mean metallicity of both the red giant star
population and the warm \ism\ in SagDIG are compared with the
properties of other dwarfs in the metallicity--luminosity diagram.
Figure~\ref{f_mateo} shows a plot of both stellar [Fe/H] and nebular
[O/H] measurements (see Skillman et al. \cite{skh89})
against luminosity for a large sample of dwarf
galaxies. The SagDIG data point was plotted using our estimates of
mean metallicity and the $M_V$ measured by KAM99.  The data are mostly
from Mateo (\cite{mateo98}) and van den Bergh (\cite{vand00}).
SagDIG now appears to be consistent with the general trend of
metallicity against luminosity for dwarf irregulars. Its metallicity,
although less extreme than found by previous studies, remains the
lowest among Local Group galaxies, and shares the tendency of
low-luminosity {\di}s (e.g., UKS\,2323-326: Lee \& Byun
\cite{mglee+byun99}; WLM: Minniti \& Zijlstra \cite{minniti97};
Sextans\,A: van Dyk et al. \cite{vandyk98})
to a lower stellar metallicity than dSph galaxies of comparable
luminosities. Whether this trend is really produced by a bimodal
metallicity, or rather is due to the fact that \rgb\ stars may have a
younger age in {\di}s than in dwarf spheroidals, is still matter of
debate.

\section{Summary and conclusions}
\label{s_summary}

We have presented a $BVI$ study of the stellar content and metallicity
of UKS~1927-177, the Sagittarius dwarf irregular galaxy, based on deep
color-magnitude diagrams. 
% Our study indicates that SagDIG is a
% ``normal'', though very metal-poor, \di\ galaxy with metallicity
% consistent with its luminosity.

We have found that {\it the color-magnitude diagram of
SagDIG is better understood by introducing a differential reddening
scenario} in which the young stars, located near high-density \hi\
clumps, are more reddened than the older stars distributed all over
the galaxy.  In fact, the color distribution of the Galactic
foreground stars indicates a relatively low reddening (we assume
$E_{B-V}=0.07 \pm 0.02$ for the old stars), while measurements of the
Balmer decrement in \hii\ regions (Skillman et al. \cite{stm89};
Saviane et al. \cite{savi+01}) provide a higher reddening,
$E_{B-V}=0.19$. 
Using the lower reddening for the ``old'' population in SagDIG, we
obtained revised values for the metallicity and distance, [Fe/H]$=-2.1
\pm 0.2$ and $(m-M)_{0}=25.14 \pm 0.18$, respectively.  While the
distance confirms previous estimates, {\it our metallicity turns out
to be significantly higher} than those proposed by Karachentsev et
al. (\cite{kara99}) and Lee \& Kim (\cite{lee00}).
{\it Using this new metallicity, we have compared SagDIG with other
dwarf galaxies in the luminosity-metallicity diagram, and found that
it is consistent with the general trend for \di\ galaxies.}

The large baseline given by the $(B-I)$ color, together with an
analysis based on statistical subtraction of the Galactic foreground,
provided additional information on the recent star formation in
SagDIG.  The C stars from two sources have also been compared with our
cleaned CMDs, which confirmed the presence of an intermediate age
population.
For the young stellar population, we have adopted the reddening
derived for the \hii\ regions, thus obtaining a good match to the
theoretical isochrones. A comparison with models indicates a
significant burst taking place $\sim$30 Myr ago, with star formation
going on until $\sim$10 Myr ago. This burst seems to have interrupted
a relatively quiescent period between $\sim$30--100 Myr ago. We
identified a group of candidate red supergiants that are quite well
fitted by the isochrones of young He-burning stars.

We have also obtained a detailed quantitative comparison of {\it the
spatial distribution of stars in different age ranges} with the
distribution of the ISM. Different {density profile scale lengths}
have been measured for the young blue stars and the red giants.
The youngest stars have been shown to concentrate in a
central region nearly coincident with the density peaks in the \hi\
distribution. 
In constrast, the distribution of red giants is quite extended, yet it
seems to be correlated with the HI morphology as well. This may
suggest for the red giants in SagDIG a contribution by 
relatively young (a few Gyr old) stars.
We also noted that some blue stars are found in ``tails'', perhaps a
hint of spiral structures, quite far from the main star-forming
sites. This suggests a similarity with the small spiral-like
structures that can form even in a small galaxy according to the
theory of stochastic self-propagating star formation (Gerola \& Seiden
\cite{gerola78}).  An alternative possibility is that these features
are a vestige of star formation related to the gaseous shell.

%--------------------------------------------------------------------
\acknowledgements We thank F. Bresolin for useful comments on an early
draft, and the anonymous referee for helpful remarks that improved the
presentation of this paper. This research has made use of the
NASA/IPAC Extragalactic Database (NED) and of NASA's Astrophysics Data
System Service.

%-----------------------------------------------------------------------------

\end{document}